\documentclass[draft,jgrga]{agutex}

\usepackage{graphicx}
\usepackage{setspace}
\usepackage{xcolor,colortbl}
\usepackage{booktabs}
\usepackage{multirow}
\usepackage{float}

 \usepackage{epsfig}

\usepackage{amssymb}
\usepackage{amsmath}
\usepackage{lineno}
\usepackage[T1]{fontenc}
\usepackage[latin1]{inputenc}
\usepackage{color}
\usepackage{pgf}

\authorrunninghead{KLAUSNER ET AL.}

\titlerunninghead{AIRGLOW AND MAGNETIC FIELD DISTURBANCES}

\authoraddr{Corresponding author: V. Klausner,
Universidade do Vale do Para\'iba,
Campus Urbanova,
Av. Shishima Hifumi, 2911- Urbanova ,
CEP 12244-000, S\~ao Jos\'e dos Campos, SP, Brazil.
(viklausner@gmail.com)}

\begin{document}


%
%

\title{Airglow and magnetic field disturbances over Brazilian region during Chile tsunami (2015)}
%
%

%
%

\authors{V. Klausner \altaffilmark{1},
F.~C.~de Meneses \altaffilmark{2},
C.~M.~N.~Candido \altaffilmark{2},
J.~R.~Abalde \altaffilmark{1}, 
P.~R.~Fagundes \altaffilmark{1}, and
E.~A.~Kherani \altaffilmark{2}
}

\altaffiltext{1}{Grupo de F\'isica e Astronomia,
Universidade do Vale do Para\'iba (UNIVAP), S\~ao Jos\'e dos Campos, SP, Brazil.}

\altaffiltext{2}{Divis\~ao de Aeronomia - DAE/CEA,
Instituto Nacional de Pesquisas Espaciais (INPE), S\~ao Jos\'e dos Campos, SP, Brazil.}









%
%


\begin{abstract}
In this work, we present first report on disturbances over Brazilian atmosphere on 16--17 September, 2015 following the Chile tsunami occurrence.
Using all-sky imager and magnetometer located at 2330 km away from the epicenter,
the presence of disturbances is noted 1--3 hours after the tsunami beginning time and during time 
which seismic tremor was also felt in the region. 
We argue that their presence towards continent at 2000-3000 km away from the epicenter offers another example of similar atmospheric response as those observed during Tohoku-Oki tsunami, 2011.
This similarity and their appearance during seismic tremor over the region classify them to be of tsunamigenic and/or seismogenic nature.
\end{abstract}

%
%

%

\begin{article}

%
%

\section{Introduction}
\label{Introduction}

Geospace monitoring of seismogenic disturbances using non-seismic sensors has gain significant momentum in the last decade.
During giant tsunamis that occurred in last ten years, tsunamigenic disturbances in the atmospheric airglow from all-sky imagers, ionospheric electron density
from GNSS receivers and geomagnetic disturbances from magnetometers were reported 
\citep{Rolland2010,Rolland2011,Makela2011,Galvanetal:2012,Kheranietall2012,Astafyeva2013,Occhipinti2013,Klausneretal:2014,Coisson2015}.

For the first time, airglow disturbances over Hawaii were observed during Tohoku-Oki tsunami \citep{Makela2011}.
An interesting aspect observed during this tsunami was the presence of tsunamigenic total electron content (TEC) disturbances towards continent opposite to the
tsunami propagation direction, covering up to 25$^\circ$ from epicentral distance \citep{Tsugawa2011,Galvanetal:2012}.
These backward propagating disturbances were simulated and interpreted as owing to the strong atmosphere shaking from the tsunami forcing near the coast \citep{Kheranietall2012}.

\begin{figure}[htb]
	\centering
		\includegraphics[width=10cm]{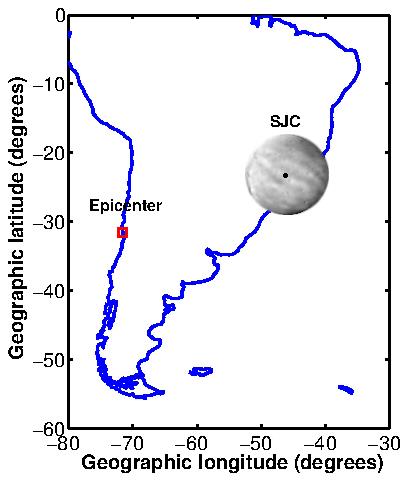}\\
		\label{fig:TTT}
	\caption{Locations of S\~ao Jos\'e dos Campos observatory (SJC -- 23.2$^\circ$S, 45.9$^\circ$W) used to geomagnetic field variations and airglow disturbances. The red square shows
the epicenter of the Chilean earthquake of 16th September, 2015. The circle indicates the field of view at a zenith angle of 90$^\circ$ projected at a $100$ km altitude
and presents a raw image for OI 557.7-nm emission obtained at 23:05 LT over the field of view. The black dot shows SJC location.}
\label{fig:map}
\end{figure}

Motivated by the presence of backward propagating tsunamigenic disturbances during Tohoku-Oki tsunami,
we search for such disturbances in the present study during the Chile tsunami that occurred on 16th
of September, 2015 at 22:55 UT, with epicenter located at $31.57^{\circ}$ S and $71.65^{\circ}$ W in $25$ km depth, and accompanied by an earthquake of magnitude $8.3$.
We select a Brazilian observatory located at S\~ao Jos\'e dos Campos (SJC - 23.2$^\circ$S, 45.9$^\circ$W) and present the observation from the all-sky imager and magnetometer.
The all-sky is a 180$^\circ$ field of view multispectral imaging system that uses four 4-inch diameter interference filter of 2.0 nm bandwidth for OI 557.7 nm emission.
The OI 557.7-nm greenline emission has been largely used to investigate wave dynamical processes around the mesopause region. 
In this kind of images, it is possible to observe gravity waves and tides signatures as well as infer the atomic oxygen profile \citep{Schubert1999}.
The OI 557.7-nm emission volumetric peak rate occurs around 95$\pm$2 km.
A fluxgate magnetometer with 0.1 nT resolution belonging to ``Universidade do Vale do Para\'iba'' is used in the present study.
This magnetometer also belongs to EMBRACE Magnetometer Network in South America.
In Figure~\ref{fig:map}, the epicenter of earthquake and SJC observatory location are shown.
Also, it is shown an airglow image over SJC with effective field of view.

\section{Results and Discussion}
\label{Results and analyses}

\subsection{Airglow Disturbances}

\begin{figure*}[htb] 
\noindent
\centering
\begin{tabular}{c c c}
\includegraphics[width=5cm]{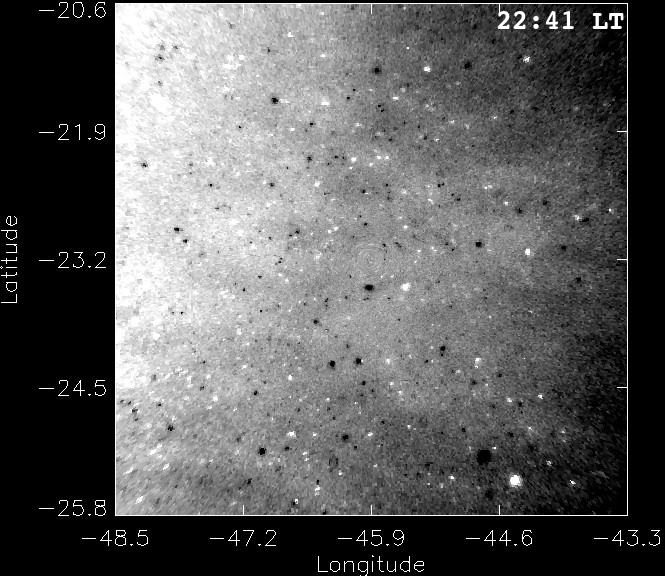}&
\includegraphics[width=5cm]{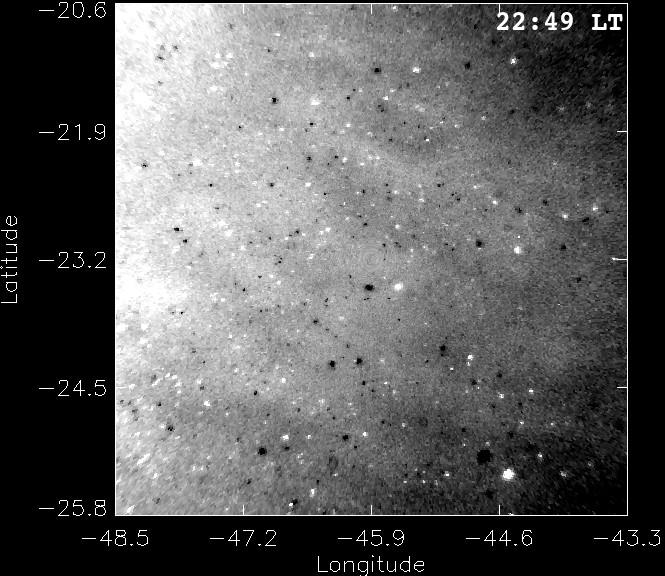}&
\includegraphics[width=5cm]{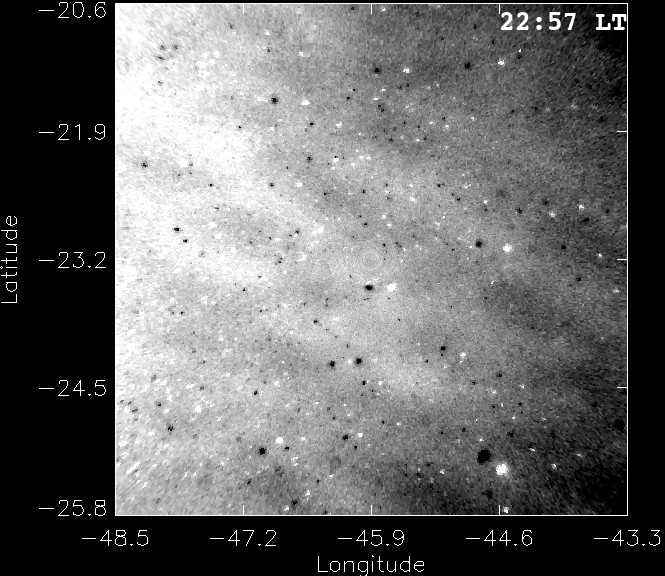}\\
\includegraphics[width=5cm]{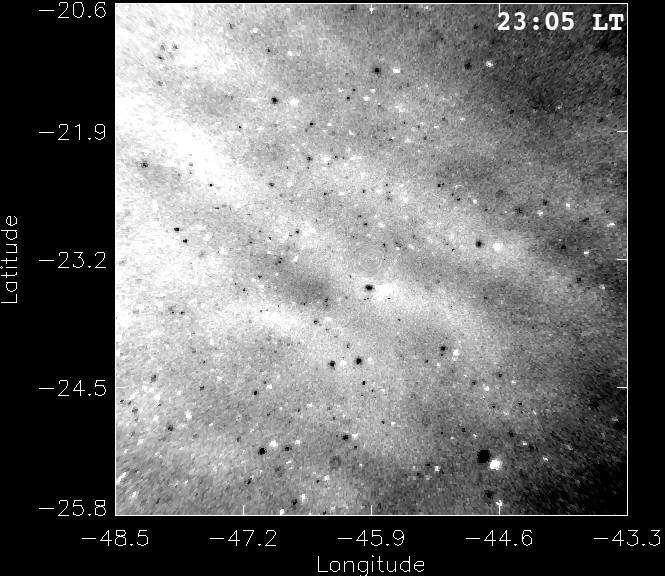}&
\includegraphics[width=5cm]{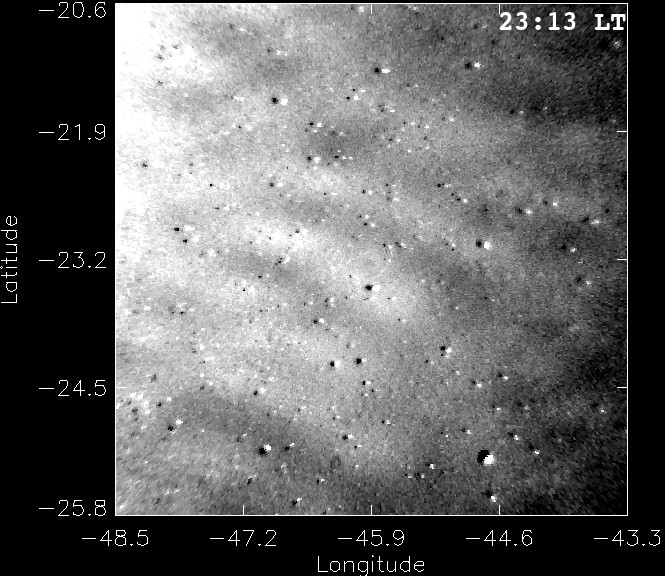}&
\includegraphics[width=5cm]{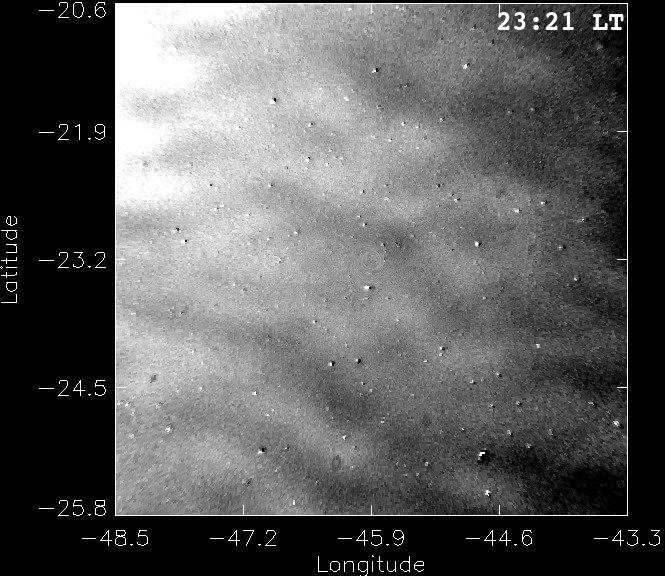}\\
\includegraphics[width=5cm]{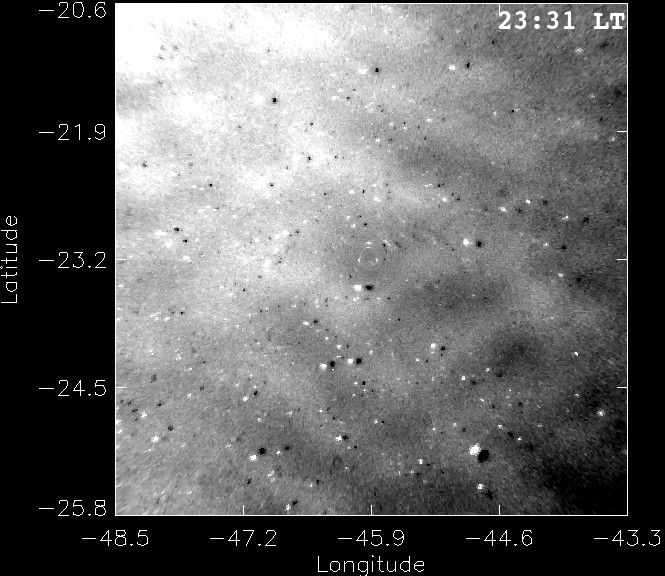}&
\includegraphics[width=5cm]{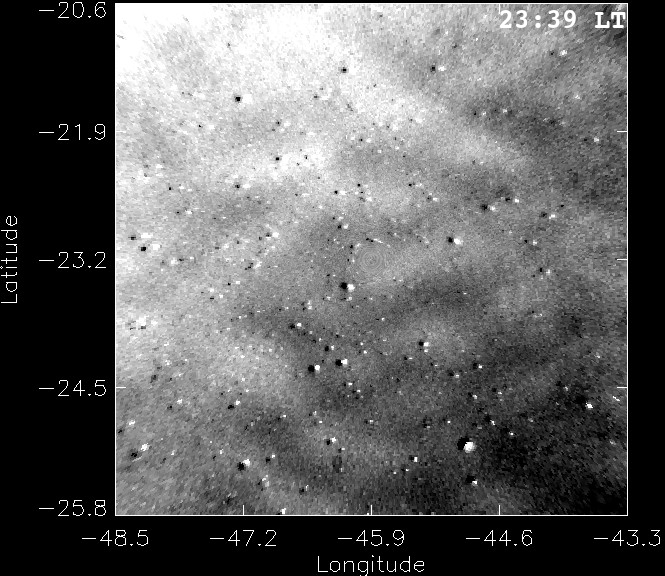}&
\includegraphics[width=5cm]{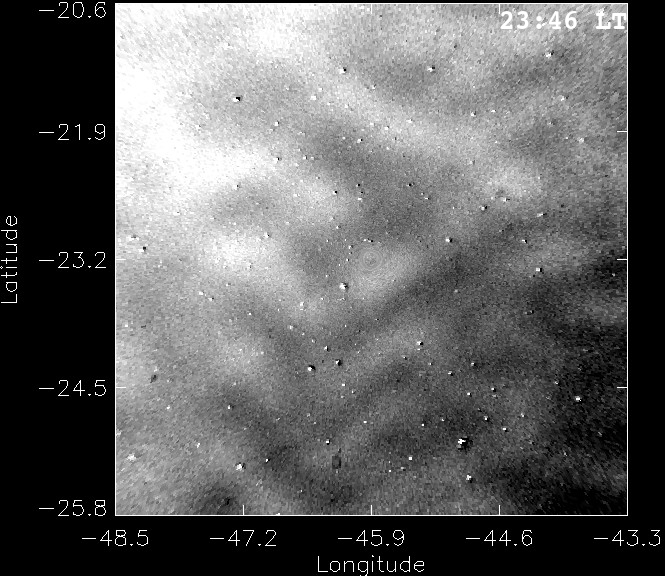}\\
\includegraphics[width=5cm]{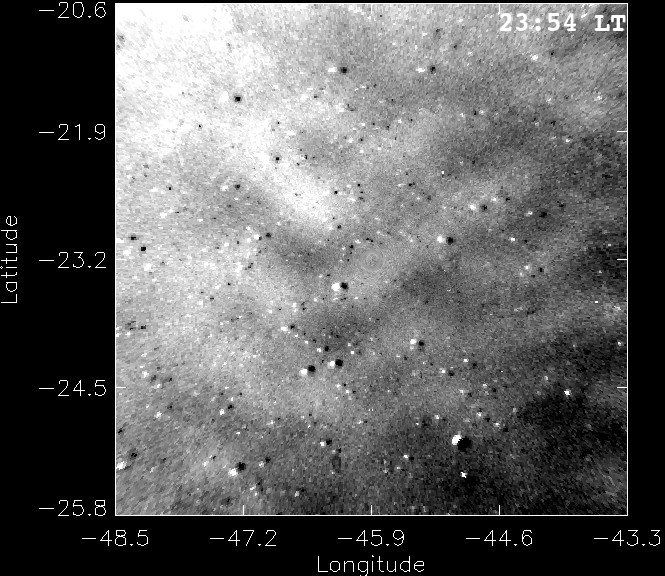}&
\includegraphics[width=5cm]{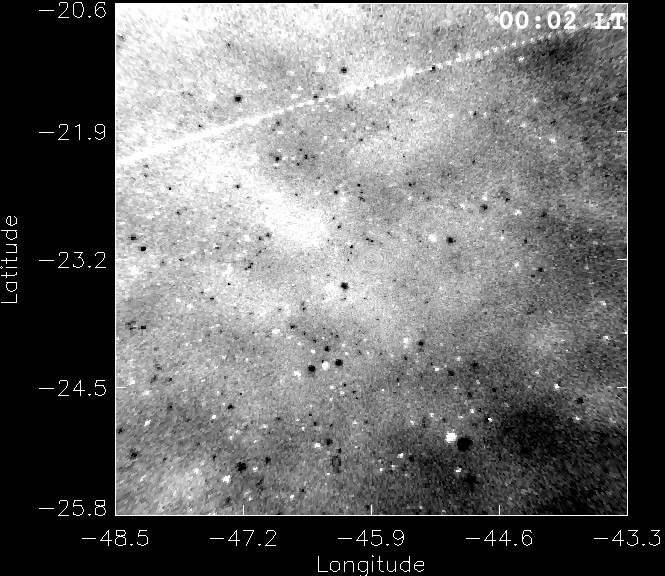}&
\includegraphics[width=5cm]{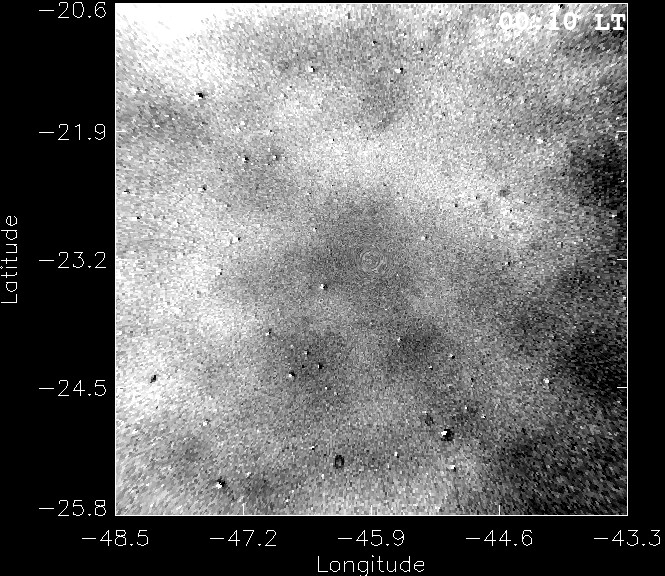}\\
\end{tabular}
\caption{\small{Sequence of the OI 557.7-nm emission all-sky images observed at ``Universidade do Vale do Para\'iba'' (UNIVAP) at S\~ao Jos\'e dos Campos (23.2$^\circ$S, 45.9$^\circ$W;),
showing the time evolution (between 22:41 LT (01:41 UT) and 00:10 LT (03:10 UT)) and spatial characteristics of gravity wave propagation on the night of 16--17th September, 2015.
The top of every imaging is located at the North geographic and left at the West geographic.}}
\label{fig:557}
\end{figure*}

\begin{figure}[htb]
\noindent
\centering\includegraphics[width=10cm]{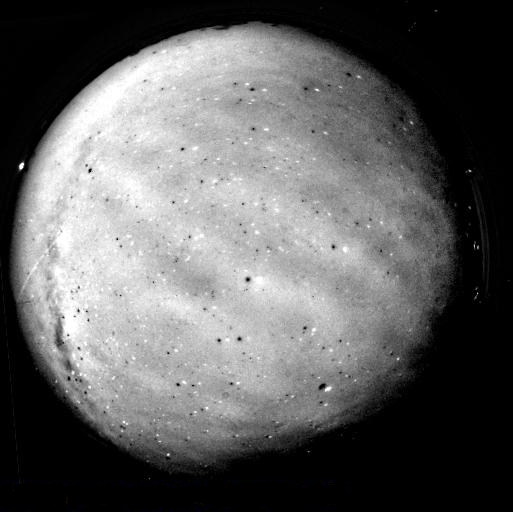}\\
 \includegraphics[width=10cm]{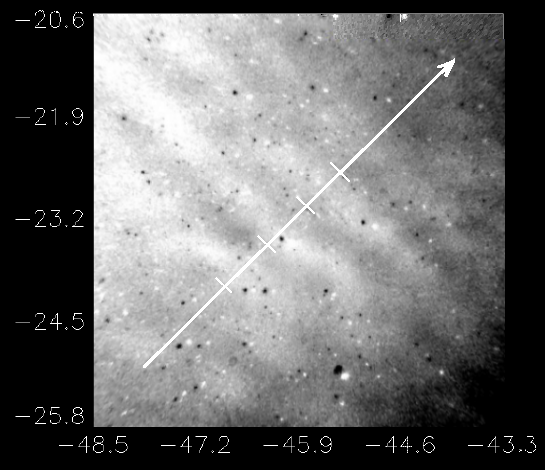}\\
 \caption{Airglow disturbance fronts propagating in Southeast direction. Vertical and horizontal axis are the latitude and longitude in degrees, respectively. The raw image at OI OI 557.7-nm at top
 and linearized at bottom obtained at 23:05 LT (02:05 UT).}
 \label{fig:direction}
\end{figure}

In Figures~\ref{fig:557}--\ref{fig:direction}, we present the images on the night of September 16--17, 2015 obtained using all-sky imager.
Figure~\ref{fig:557} shows sequences of images of the OI 557.7-nm emission in which we note the presence of faint band-like airglow disturbances that
appear around 02:05 UT (3 hours and 5 minutes after the earthquake event) and last about 30 minutes.
These disturbances have wavefront aligned in northwest-southeast and propagated in northeastward direction, \textit{i. e.}, in the opposite direction to the tsunami propagation.
From an enlarged image at 02:05 UT in Figure~\ref{fig:direction}, we note that these disturbances have wavelength in between 60 to 100 km.
The airglow movie for the night of this reported event and the night before and after are available in the auxiliary material as Movie\_15Sep2015, Movie\_16Sep2015 and Movie\_17Sep2015 for comparison.
From these movies, we verify that such disturbances as noted on the night of tsunami are absent on other days.

Band-like airglow disturbances are common feature over Brazilian region, and they are associated to gravity waves/medium-scale 
traveling ionospheric disturbances (GWs/MSTIDs) \citep{Medeiros2003,Wrasse2006,Candidoetal2008,Pimenta2008}. 
Both mesospheric GWs and MSTIDs signatures observed in airglow images by \cite{Medeiros2003,Wrasse2006,Candidoetal2008,Pimenta2008}
are commonly attributed to tropospheric convective processes or to middle latitude instabilities.
In addition, they have opposite orientation (northeast to southwest) and propagation direction (northwestward) to those observed here.
Therefore, the observed airglow disturbances in Figures~\ref{fig:557}--\ref{fig:direction} are not the convectively-driven GWs/MSTIDs commonly observed in the region.

On the other hand, these disturbances are possibly seismogenic or tsunamigenic.
For the Tohoku-Oki tsunami, ionospheric disturbances in TEC were reported to propagate towards backward direction in the form of concentric wavefronts, and
to arrive up to 23$^\circ$ from epicentral distance \citep{Tsugawa2011,Galvanetal:2012}.
The airglow disturbances in Figures~\ref{fig:557}--\ref{fig:direction} offer a similar scenario where their presences are noted at 2000--3000 km ($\sim$18$^\circ$ -- 27$^\circ$) from epicentral distance
in the backward direction.

For the Tohoku-Oki tsunami, the backward propagating TEC disturbances were simulated and interpreted as owing to the strong atmosphere shaking from the tsunami forcing near the coast \citep{Kheranietall2012}.
Therefore, the observed backward propagating airglow disturbances in the present study are possibly arising from the similar atmospheric shaking.

\subsection{Geomagnetic Field Disturbances}

\begin{figure}[htb]
\noindent
\centering
 \includegraphics[width=14cm]{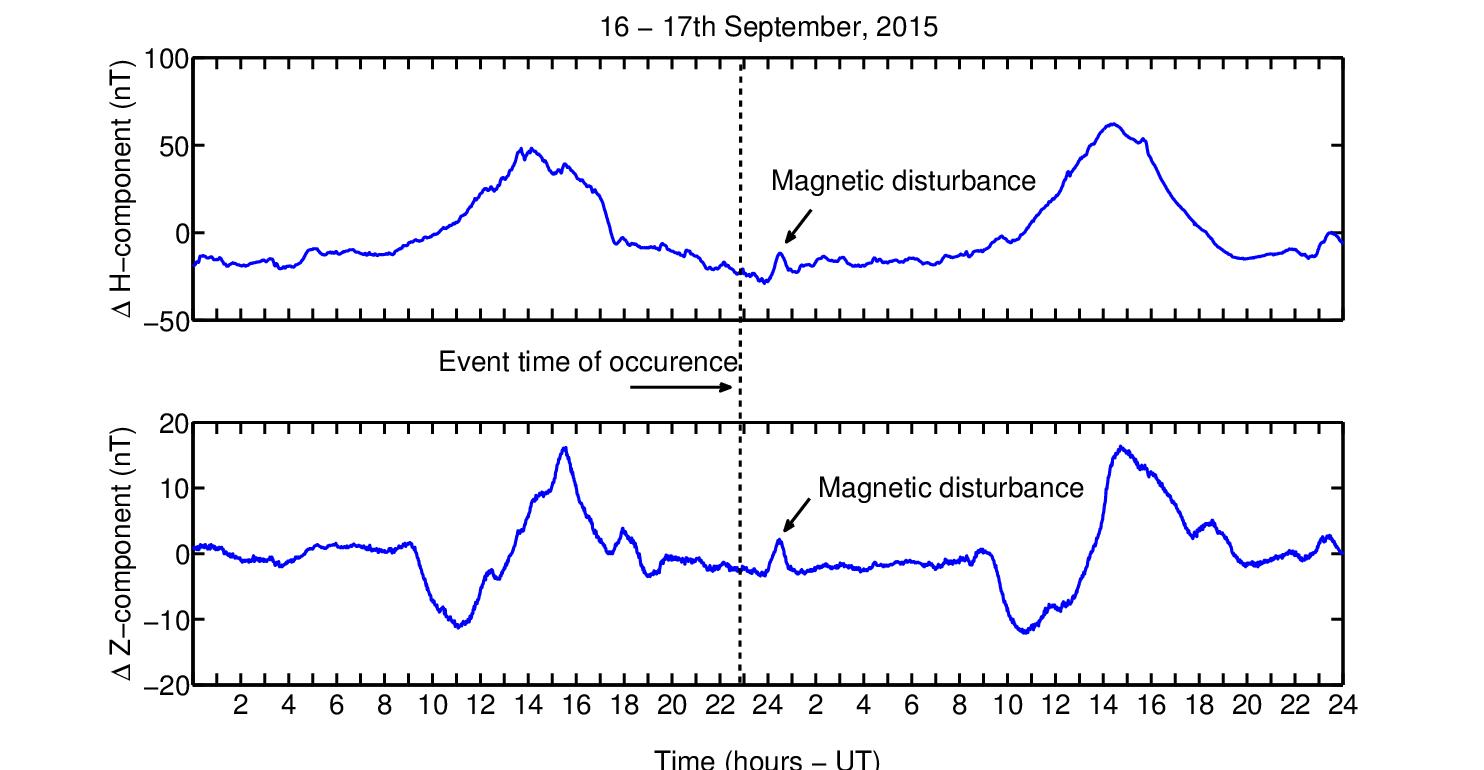}\\
 \caption{Minutely magnetogram data from the H-component (top) and Z-component (bottom) variations over S\~ao Jos\'e dos Campos. The dashed line shows the time of the Chilean earthquake (2015) and
 the arrows point to the increase of the geomagnetic field over S\~ao Jos\'e dos Campos several minutes after of the event occurrence.}
 \label{fig:event}
\end{figure}

\begin{figure*}[htb]
\noindent
\centering
 \includegraphics[width=14cm]{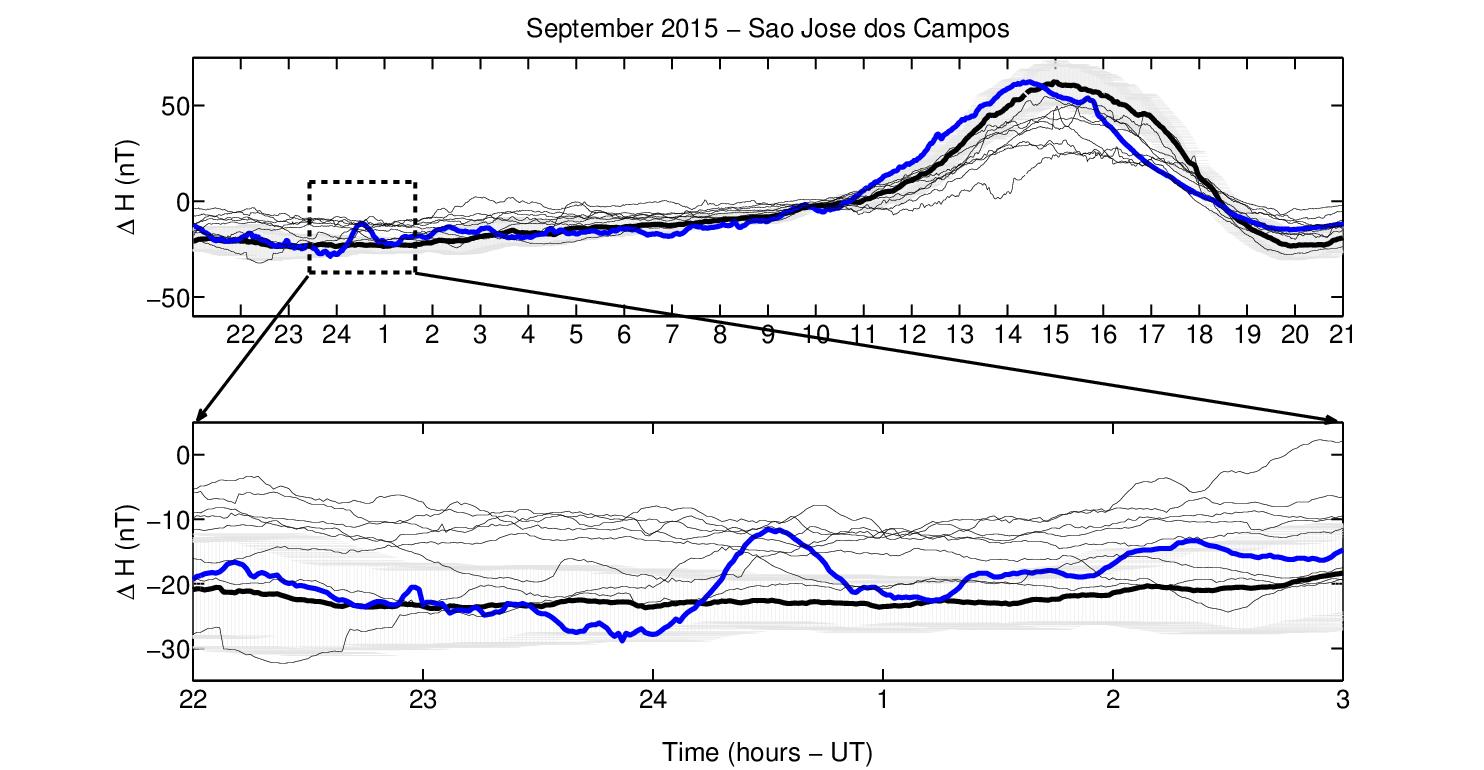}\\
  \includegraphics[width=14cm]{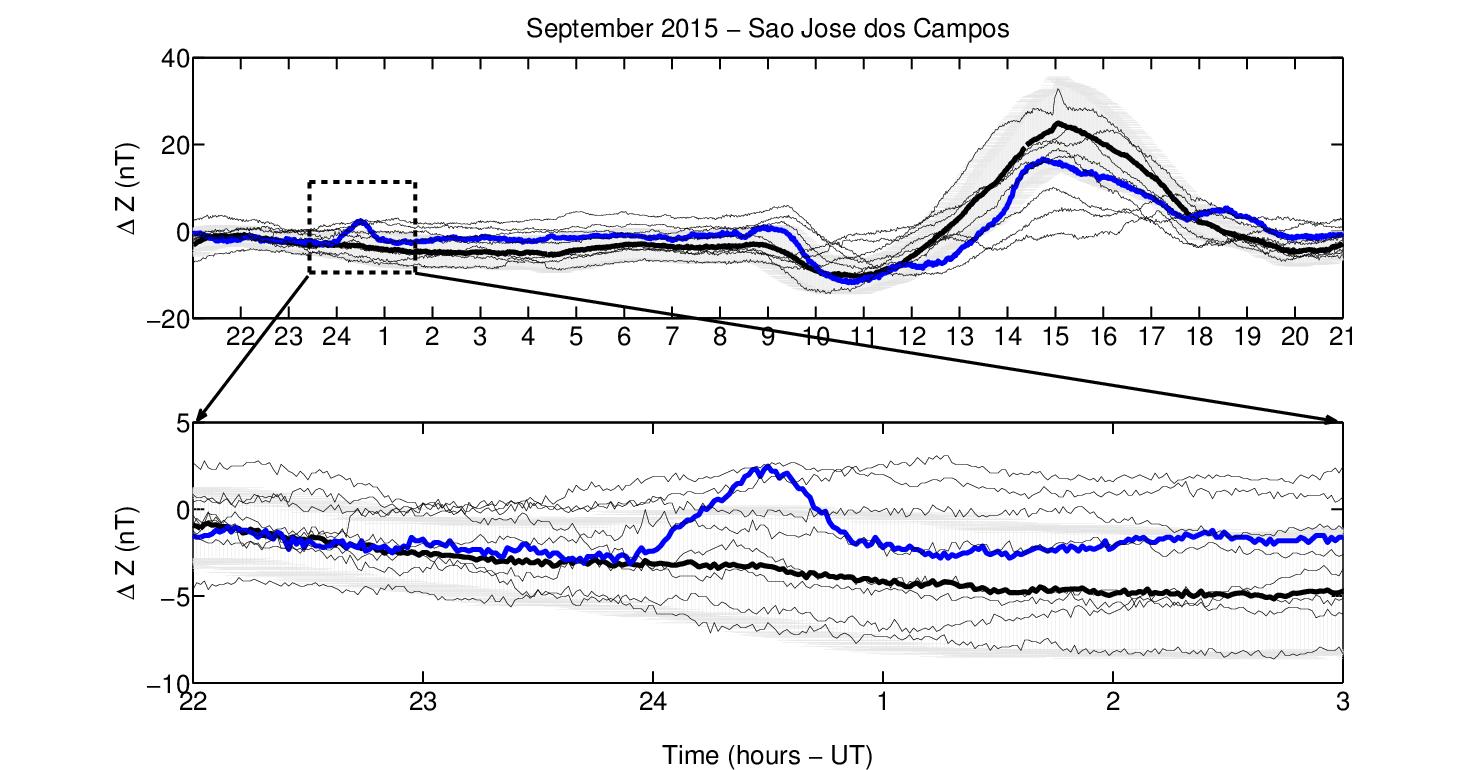}
  \caption{Minutely magnetogram data for days of September with Dst$\geq$-20 nT (Aug.31/1st--2/3th, 24/25th--28/29th), 2015 (black thin lines). 
  From top to bottom, each panel shows the H-component and Z-component variations over S\~ao Jos\'e dos Campos. 
  Enlargement shows a magnetic pulse (blue color) at S\~ao Jos\'e dos Campos about 40 minutes after the earthquake.
  The black thick line corresponds to mean value variation of these all eight quietest days and the blue color line corresponds to the day of 16/17th September,2015.
  On the horizontal axis, the time varies from 21:01 UT (day before) until 21:00 UT (24 hours period).}
  \label{fig:compHeZ}
\end{figure*}

On Figures~\ref{fig:event}--\ref{fig:compHeZ}, geomagnetic field measurements are presented.
In Figure~\ref{fig:event}, the H- and Z-components during 16-17 September, 2015 are plotted.
In Figure~\ref{fig:compHeZ}, the mean value (black color) of the quietest day of September (2015) and the standard deviation
(grey color) between each day to the mean value are plotted.
The 16th of September, 2015 is highlighted in blue color.

In Figure~\ref{fig:event}, we note a magnetic pulse around 24:00 UT, about 40 minutes after the earthquake.
Such pulse is not present on previous night.
Also, we note the presence of such pulse in Figure~\ref{fig:compHeZ} on night of tsunami while no such pulse is present in any other quiet days of September, 2015.
Moreover, this pulse appears during the time when the seismic tremors were felt in the region.
Consequently, these aspects provide further evidence that this magnetic pulse is seismogenic or tsunamigenic.

\section{Conclusions}
\label{Conclusions} 

We document first report on the airglow and geomagnetic field disturbances related to the recently occurred Chile tsunami (2015).
Also, these disturbances are observed over the Brazilian sector for the first time.
The fact of airglow disturbances could be detected in opposite direction of the tsunami propagating is very encouraging, and
it shows that the constant monitoring of the ionosphere and geomagnetic field could play an important role to calibrate seismic/tsunami models,
and used in understanding of the physical processes involved in the tsunami propagation.


%
%

%
%
%
%
%

\begin{acknowledgments}
V. Klausner and C. M. N. Candido wish to thanks their Postdoctoral research for the financial support within the Programa Nacional de P\'os-Doutorado (PNPD -- CAPES) at UNIVAP and INPE, respectively.
We wish to express his sincere thanks to the Funda\c{c}\~ao de Amparo a Pesquisa do Estado de S\~ao Paulo (FAPESP), for providing financial support through the process numbers 2009/15769-2 and 2012/08445-9,
CNPq grants number 457129/2012-3, 
and FINEP number 01.100661-00 for the partial financial support.
\end{acknowledgments}

\end{article}

\newpage


%
%

%

%

%
%
%
%
%
%



\begin{thebibliography}{}

\bibitem[{\textit{Astafyeva et al.}(2013)\textit{Astafyeva, Shalimov, Onshanshaya, and Lognonn\'e}}]{Astafyeva2013} 
Astafyeva, E., S.~Shalimov, E.~Onshanshaya, and P.~Lognonn\'e (2013)
Ionospheric response to earthquakes of different magnitudes: Larger quakes perturb the ionosphere stronger and longer
\textit{Geophysical Research Letters}, \textit{40}~(9), 1675--1681.

\bibitem[{\textit{Candido et al.}(2008)\textit{Candido, Pimenta, Bittencourt, and Becker-Guedes}}]{Candidoetal2008}
Candido, C. M. N., A.~A.~Pimenta, J.~A.~Bittencourt, and F.~Becker-Guedes (2008),
Statistical analysis of the occurrence of medium-scale traveling ionospheric disturbances over Brazilian low latitudes using OI 630.0 nm emission all-sky images, 
\textit{Geophysical Research Letters}, \textit{35}, L17105. 

\bibitem[{\textit{Coisson et al.}(2015)\textit{Coisson, Lognonn\'e, Walwer, and Rolland}}]{Coisson2015}
Coisson, P., P.~Lognonn\'e, D.~Walwer, and L.~M.~Rolland (2015),
First tsunami gravity wave detection in ionospheric radio occultation data, 
\textit{Earth and Space Science}, \textit{2}, 125--133.

\bibitem[{\textit{Galvan et~al.}(2012)\textit{Galvan, Komjathy, Hickey, Stephens, Snively, Song, Butala, and Mannucci}}]{Galvanetal:2012}
Galvan, D.~A., A., Komjathy, M.~P., Hickey, P., Stephens, J., Snively, Y.~T., Song, M.~D., Butala, and A.~J., Mannucci (2012),
Ionospheric signatures of Tohoku-Oki tsunami of March 11, 2011: Model comparisons near the epicenter.
\textit{Radio Science}, \textit{47}(4), RS4003.

\bibitem[{\textit{Kherani et~al.}(2012)\textit{Kherani, Lognonn\'e, H\'ebert, Rolland, Astafyeva, Occhipinti, Co\"isson, Walwer, and de Paula}}]{Kheranietall2012}
Kherani, E.~A., P.~Lognonn\'e, H.~H\'ebert, L.~Rolland, E.~Astafyeva, G.~Occhipinti, P.~Co\"isson, D.~Walwer, E.~R.~de Paula (2012),
Modelling of the total electronic content and magnetic field
anomalies generated by the 2011 Tohoku-Oki tsunami and associated acoustic-gravity waves,
\textit{Geophysical Journal International}, \textit{191}(3), 1049--1066.

\bibitem[{\textit{Klausner et~al.}(2014)\textit{Klausner, Mendes, Domingues, Papa, Tyler, Frick, and Kherani}}]{Klausneretal:2014}
Klausner, V., O., Mendes, M.~O., Domingues, A.~R.~R. Papa, R.~H., Tyler, P., Frick, and E.~A. Kherani (2014),
Advantage of wavelet technique to highlight the observed geomagnetic perturbations linked to the Chilean tsunami (2010)
\textit{Journal of Geophysical Research: Space Physics}, \textit{119}(4), 3077--3093.

\bibitem[{\textit{Makela et~al.}(2011)\textit{Makela, Lognonn\'e, H\'ebert, Gehrels, Rolland, Allgeyer, Kherani, Occhipinti, Astafyeva, Coisson, Loevenbruck, 
Cl\'ev\'ed\'e, Kelley, and Lamouroux}}]{Makela2011}
Makela, J. J., P.~Lognonn\'e, H.~H\'ebert, T.~Gehrels, L.~Rolland, S.~Allgeyer, A.~E.~Kherani, G.~Occhipinti, E.~Astafyeva, P.~Coisson, A.~Loevenbruck, 
E.~Cl\'ev\'ed\'e, M.~C.~Kelley, and J.~Lamouroux (2011),
Imaging and modeling the ionospheric airglow response over Hawaii to the tsunami generated by the Tohoku earthquake of 11 March 2011,
\textit{Geophysical Research Letters}, \textit{38}, L00G02.

\bibitem[{\textit{Medeiros et al.}(2003)\textit{Medeiros, Taylor, Takahashi, Batista, and Gobbi}}]{Medeiros2003}
Medeiros, A. F.,  M.~J.~Taylor,  H.~Takahashi, P.~P.~Batista, and D.~Gobbi (2003),
An investigation of gravity wave activity in the low-latitude upper mesosphere: Propagation direction and wind filtering,
\textit{Journal of Geophysical Research}, \textit{108}(D14), 4411.

\bibitem[{\textit{Occhipinti et~al.}(2013)\textit{Occhipinti, Rolland, Lognonne, and Watada}}]{Occhipinti2013}
Occhipinti, G., L. Rolland, P. Lognonn\'e, S. Watada (2013), 
From Sumatra 2004 to Tohoku-Oki 2011: The systematic GPS detection of the ionospheric signature induced by tsunamigenic earthquakes.
\textit{Journal of Geophysical Research: Space Physics}, \textit{118}(6), 3626--3636.

\bibitem[{\textit{Pimenta et~al.}(2008)\textit{Pimenta, Amorim, and Candido}}]{Pimenta2008} 
Pimenta, A. A., D.~C.~M.~Amorim, and C.~M.~N.~Candido (2008), 
Thermospheric dark band structures at low latitudes in the Southern Hemisphere under different solar activity conditions: A study using OI 630 nm emission all-sky images, 
\textit{Geophysical Research Letters}, \textit{35}, L16103. 

\bibitem[{\textit{Rolland et~al.}(2010)\textit{Rolland, Occhipinti, Lognonne, and Loevenbruck}}]{Rolland2010}
Rolland, L., G.~Occhipinti, P.~Lognonn\'e, and A.~Loevenbruck (2010), 
Ionospheric gravity waves detected offshore Hawaii after tsunamis.
\textit{Geophysical Research Letters}, \textit{37}(17), L17101.

\bibitem[{\textit{Rolland et~al.}(2011)\textit{Rolland, Lognonne, and Munekane}}]{Rolland2011}
Rolland, L., P.~Lognonn\'e, and H.~Munekane (2011), 
Detection and modeling of Rayleigh wave induced patterns in the ionosphere.
\textit{Geophysical Research Letters}, \textit{116}, A05320.

\bibitem[{\textit{Schubert et~al.}(2011)\textit{Schubert, Walterscheid, Hickey, and Tepley}}]{Schubert1999}
Schubert, G., R.~L.~Walterscheid, M.~P.~Hickey, and C.~A.~Tepley (1999),
Observations and interpretations od gravity-wave induced fluctuations in the OI (557.7-nm) airglow,
\textit{Journal of Geophysical Research}, \textit{104}~(A7),14915--14924.

\bibitem[{\textit{Tsugawa et al.}(2011)\textit{Tsugawa, Saito, Otsuka, Nishioka, Maruyama, Kato, Nagatsuma, and Murata}}]{Tsugawa2011} 
Tsugawa, T., A.~Saito, Y.~Otsuka, M.~Nishioka, T.~Maruyama, H.~Kato, T.~Nagatsuma, and K.~T.~Murata (2011),
Ionospheric disturbances detected by GPS total electron content observation after the 2011 off the Pacific coast of Tohoku Earthquake
\textit{Earth, Planets and Space}, \textit{63}~(7), 875--879.
    
\bibitem[{\textit{Utada et~al.}(2011)\textit{Utada, Shimizu, Ogawa, Maeda, Furumura, Yamamoto, Yamazaki, Yoshitake and Nagamachi}}]{Utadaetal:2011}
Utada, H., H.~Shimizu, T.~Ogawa, T.~Maeda, T.~Furumura, T.~Yamamoto, N.~Yamazaki, Y.~Yoshitake and S.~Nagamachi (2011),
Geomagnetic field changes in response to the 2011 off the Pacific Coast of Tohoku Earthquake and Tsunami,
\textit{Earth and Planetary Science Letters}, \textit{311}(1--2), 11--27.

\bibitem[{\textit{Zhang et~al.}(2014)\textit{Zhang, Baba, Liang, Shimizu, and Utada}}]{Zhangetal:2014}
Zhang, L., K.~Baba, P.~Liang, H.~Shimizu, and H.~Utada (2014),
The 2011 Tohoku tsunami observed by an array of ocean bottom electromagnetometers, 
\textit{Geophysical Research Letters}, \textit{41}.

\bibitem[{\textit{Wrasse et~al.}(2006)\textit{Wrasse, Nakamura, Takahashi, Medeiros, Taylor, Gobbi, Denardini, Fechine, Buriti, Salatun, Suratno, Achmad, and Admiranto}}]{Wrasse2006}
Wrasse, C. M., T.~Nakamura, H.~Takahashi, A.~F.~Medeiros, M.~J.~Taylor, D.~Gobbi, C.~M.~Denardini, J.~Fechine, R.~A.~Buriti, A.~Salatun, Suratno, E.~Achmad, and A.~G.~Admiranto (2006),
Mesospheric gravity waves observed near equatorial and low-middle latitude stations: wave characteristics and reverse ray tracing results,
\textit{Annales Geophysicae}, \textit{24}, 3229--3240.

%
%
%
%

\end{thebibliography}
\end{document}